\documentstyle[12pt,a4]{article}
\begin{document}
\begin{center}
\large {\bf Quantum Zeno Effect and Light-Dark Periods \\ for a Single Atom}
\vskip.5cm
Almut Beige\footnote
{e-mail: beige@theorie.physik.uni-goettingen.de}
and Gerhard C. Hegerfeldt\footnote
{e-mail: hegerf@theorie.physik.uni-goettingen.de}\\[.5cm]
\normalsize
   Institut f\"ur Theoretische Physik\\
   Universit\"at G\"ottingen\\
   Bunsenstr.~9\\
   D-37073 G\"ottingen, Germany
\end{center}
\begin{abstract}
The quantum Zeno effect (QZE) predicts a slow-down of the
time development of a system under rapidly repeated ideal
measurements, and experimentally this was tested for an ensemble of
atoms using short laser pulses for non-selective state measurements.
Here we consider such pulses for selective measurements on a {\em single}
system. Each probe pulse will cause a burst of fluorescence or no
fluorescence. If the probe pulses were strictly
ideal measurements, the QZE would predict periods of fluorescence
bursts alternating with periods of no fluorescence (light and dark
periods) which would become longer and longer with increasing
frequency of the measurements. The non-ideal character of the
measurements is taken
into account by incorporating the laser pulses in the
interaction, and this is used to determine the corrections to the
ideal case. In the limit, when the time $\Delta t$ between the laser
pulses goes to zero, no freezing occurs but instead we show
convergence to the familiar macroscopic
light and dark periods of the continuously driven Dehmelt system.
An experiment of this type should be feasible for a
single atom or ion in a trap.\\[.25cm]
PACS numbers 03.65.Bz; 42.50.-p; 32.90.+a
\end{abstract}
\vspace*{0.5cm}

\noindent {\bf 1. Introduction}
\vspace*{0.5cm}

The effect of an instantaneous measurement on a quantum mechanical
system is usually described by the projection postulate of von
Neumann and L\"uders \footnote{
The projection postulate as currently used has been
formulated by L\"uders \cite{Lue}.
For observables with degenerate eigenvalues his formulation differs
from that of von Neumann \cite{N}.
It has been pointed out to us by A. Sudbury (private communication)
that in the first edition of his book Dirac \cite{Dirac}
defines observations which cause minimal disturbance and which
correspond to L\"uder's prescription; in later editions, however, this
passage has been omitted.}
according to which, depending on the
outcome of a measurement, the wave-function of the
system is projected onto the respective eigenspaces of the observable
under consideration. This is also called reduction or collapse of the
wave-function under an ideal measurement;
a more general approach to measurements is taken in \cite{Ludw}.
Using this concept
and some fairly general technical assumptions
Misra and Sudarshan \cite{MiSu} have investigated how a system is
affected by rapidly repeated ideal measurements at times $\Delta t$ apart.
They found a slow-down of the
system's time development and, in the limit $\Delta t \rightarrow
0$, a freezing of the state. This is called the quantum Zeno effect
(QZE). The basic reason for this is the fact
that for short enough times transition probabilities grow only
quadratically with time, not linearly.

To test this effect, Itano et al.~\cite{Wine} performed an experiment
with an ensemble of 5000 ions in a trap (see Fig. 1 for the relevant
level structure, a V configuration).
The time development was given by a so-called
$\pi$ pulse of length $T_{\pi}$, tuned to the 1
- 2 transition frequency.
A $\pi$ pulse, here an rf pulse, transforms the initial
state $|1\rangle$ into $|2 \rangle $ at the end of the pulse, if no
measurements are performed.
Following a proposal of Cook \cite{Cook}
the population of the lower level was measured
{}~--~ non-selectively
and without actually recording the results ~--~ in rapid
succession through the fluorescence induced
by very short pulses of a strong probe laser
which couple level 1 with an auxiliary third level.
The population at time $T_\pi$ was
then measured by a final pulse and recorded.
The experimental results were in good
agreement with the predictions of the QZE.

The QZE and this experiment have
not only aroused considerable interest in the
literature \cite{interest,crit1}, but the very
relevance of the above experimental results for the QZE
has given rise to controversies. In
particular the projection postulate and its applicability
in this experiment have been cast into doubt, and
it was pointed out that the experiment could
be understood without recourse to the QZE by simply including the
probe laser in the dynamics, e.g.~in the Bloch equations or
in the Hamiltonian \cite{crit1}. Since
the Bloch equations describe the
density matrix of the {\em complete} ensemble, including the probe
pulse as an interaction
in them gives, however, no direct insight on how such a pulse acts
on a single system.

In previous papers \cite{BeHe1,BeHeSo,BeHe2} we have therefore
investigated in how
far a short laser pulse realizes a selective measurement,
i.e.~on single systems, to which the projection
postulate can be applied. By means of the quantum jump approach
(or Monte Carlo wave functions or quantum trajectories) \cite{QJ}
and including the probe laser in the dynamics
we showed analytically that for a wide range of
parameters such a short laser pulse
acts indeed
as an {\em effective} level measurement to which the usual projection
postulate applies with high accuracy. The corrections to the ideal
reductions and their accumulation over $n$ pulses were calculated.
Our conclusion was that the projection postulate
is an excellent pragmatic tool for a quick and intuitive understanding of the
slow-down of the time evolution in experiments of this type and that it gives a
good physical insight. But it is only approximate, and a more
detailed analysis has to take the corrections into account.

The experiment of Ref.~\cite{Wine} deals with the effect of repeated
non-selective measurements on an ensemble of systems and with the
associated slow-down in the time evolution of the density matrix of
the total ensemble.
It suggests itself to perform a similar experiment with a single atom
(or ion) in a trap, though not only for the duration of a $\pi$ pulse
of the weak driving field but instead for an arbitrary {\em long}
time. This might be regarded as an analog of the idealized situation
of rapidly repeated measurements on a single system. As studied in
Refs.~\cite{MiSu,Cook}, in the idealized situation the outcome of
the measurements will form a stochastic sequence, in this case a
sequence of states $|1\rangle$ and $|2\rangle$. The periods containing
only $|1\rangle$'s and $|2\rangle$'s will become increasingly long
when the time $\Delta t$ between the ideal measurements decreases,
and in the limit $\Delta t \rightarrow 0$ one would have a single
infinite sequence of $|1\rangle$'s or $|2\rangle$'s, i.e.~freezing.
With short pulses of a probe laser, considered as measurements,
one would therefore expect periods of fluorescence bursts (light
periods, corresponding to periods of $|1\rangle$'s) alternating with
periods of no fluorescence (dark periods, corresponding to periods of
$|2\rangle$'s). Decreasing the time $\Delta t$ between the probe
pulses should, in this picture, make the light and
dark periods longer.

The aim of this paper is to analyze how far this intuitive picture of
the behavior of a single system is correct and to provide an
understanding why the projection postulate also works so well in this
case. After a brief review of the ideal case we use our previous
results to calculate in Section III the mean duration of the light
and dark periods, $T_L$ and $T_D$, and compare them to the simple
expression obtained by the projection postulate. Our analysis will
make it perfectly clear why the projection postulate gives such
excellent results for a wide range of parameters. If the time $\Delta
t$ between the probe pulses becomes too small, however, then the
above simple picture breaks down. In Section IV we will explicitly
perform the limit $\Delta t \rightarrow 0$ and show that in contrast
to the idealized case $T_L$ and $T_D$ remain finite. Indeed, we
show convergence to the same expressions
as for the famous light and dark periods
of the continuously driven Dehmelt system, which are also known under
the name of `electron shelving' \cite{Dehmelt}. In the last section
we discuss our results.
\vspace*{0.5cm}

\noindent {\bf 2. Brief review of ideal case}

\vspace*{0.5cm}

If one performs rapidly repeated ideal measurements of an observable $A$
with discrete eigenvalues on a single system at times $\Delta t$ apart
then the projection postulate predicts that one will find the same
value of $A$ in a row for some time, then another value for some time,
and so on. The length of these time intervals is stochastic, and their
lengths increase when $\Delta t$ decreases. For an observable $A$ with
non-degenerate discrete eigenvalues this can be seen as follows.
For simplicity we make a domain assumption further below.
For the general treatment see Ref.~\cite{MiSu}.

Let $|a\rangle$ be a state vector and
$I\!\!P_a \equiv | a\rangle\langle a |$ the
corresponding projector. At times $t_1, t_2,...,$ with $\Delta t
\equiv t_{i + 1} - t_i$, ideal measurements of $I\!\!P_a$ are
performed, whose results are 1 or 0, with the system afterwards in
$|a\rangle$ or the subspace orthogonal to $|a\rangle$, respectively.
This is equivalent to asking whether the
result of a measurement is $|a\rangle$ or perpendicular to
$|a\rangle$, and we denote the outcome $a$ and $\bot$ instead of 1
and 0. We define $I\!\!P_\bot =1\!\!1-I\!\!P_a$.
Let $U(t, t')$ be the time-development
operator for the system. If, for initial state $| a\rangle$, one has found
$a$ in $n$ successive measurements, the resulting state is, up to
normalization, given by
\begin{equation}\label{1}
|\psi_a(t_n, t_0)\rangle \equiv I\!\!P_a U(t_n, t_{n-1}) I\!\!P_a
....I\!\!P_a U(t_1, t_0) |\psi\rangle~,
\end{equation}
which of course is proportional to $|a\rangle$,
and the probability $P_a(t_n, t_0; | \psi\rangle)$ for this is
\begin{eqnarray}\label{2}
P_a(t_n, t_0; | \psi\rangle)
&=& \| \,| \psi_a(t_n, t_0)\rangle\|^2 \nonumber \\
&=& | \langle a | U(t_1, t_0) | \psi \rangle |^2
\prod^n_{i = 2} |\langle a | U(t_i, t_{i-1}) | a\rangle |^2 ~.
\end{eqnarray}
If one has found $\bot$ in $n$ successive measurements the
state is
\begin{equation}\label{3}
|\psi_\bot (t_n, t_0)\rangle =
I\!\!P_\bot U(t_n, t_{n-1}) I\!\!P_\bot
\cdots {I\!\!P_\bot} U(t_1, t_0) | \psi\rangle~,
\end{equation}
which in general is no longer proportional to a fixed vector, and
the probability for this is given by
\begin{eqnarray}
P_\bot (t_n, t_0;|\psi\rangle) &=& \| \,|\psi_\bot (t_n,t_0)\rangle\|^2~.
\nonumber
\end{eqnarray}

To show that, for fixed $t = n \Delta t, P_a (t, t_0) \rightarrow 1
\cdot |\langle a| \psi\rangle|^2$ for $\Delta t \rightarrow 0$ we assume for
simplicity
that $|a\rangle$ is in the domain of $H$. An expansion then gives
\cite{Lukacs}
\begin{eqnarray} \label{4}
|\langle a | U(t_i, t_{i-1})|a\rangle|^2
&=& 1 - \Delta t^2 [\langle a | HH|a\rangle
- \langle a|H|a\rangle^2]/\hbar^2+ o(\Delta t^2) \nonumber\\
&=& {\rm e}^{- \Delta t^2 [\langle a|H^2|a\rangle-\langle
a|H|a\rangle^2]/\hbar^2}(1 + o(\Delta t^2))
\end{eqnarray}
where $o(\Delta t^2)$ denotes terms which go to 0 faster than
$\Delta t^2$.
The expression $\langle a|HH|a\rangle$
is to be interpreted as $||H|a\rangle||^2$.
Eq.~(\ref{4}) just states the well-known fact
that under the above assumptions the transition probability from
$|a\rangle$ to an orthogonal state goes as $\Delta t^2$ for small
$\Delta t$ \cite{Khalf}. From Eqs.~(\ref{2}) and (\ref{4}) one now
obtains for the probability
\begin{eqnarray} \label{5}
P_a(t,t_0; | \psi\rangle)
&=& {\rm e}^{-(n-1) \Delta t^2 [\langle a|H^2|a\rangle
-\langle a |H| a\rangle^2 ]/\hbar^2}
(1 + o(\Delta t^2))^{n-1} |\langle a|U(t_1, t_0)|\psi\rangle|^2~.
\end{eqnarray}
With $n = t/ \Delta t$ the first and second factor in Eq.~(\ref{5}) go
to 1 for $\Delta t \rightarrow 0$, and the last to $|\langle a | \psi\rangle
|^2$.

Under the same conditions one can also show
that $P_\bot(t, t_0; | \psi\rangle
) \rightarrow 1 \cdot \| I\!\!P_\bot | \psi\rangle \|^2$ for
$\Delta t \rightarrow 0$. If $ I\!\!P_\bot$ were a one- or
finite-dimensional projector this would follow as before, but in the
general case another argument is needed. With $U_{\Delta t} \equiv
U(\Delta t, 0)$ one has from Eq.~(\ref{3})
\begin{eqnarray} \label{6}
P_\bot (t_i, t_0; | \psi\rangle)
- P_\bot(t_{i+1}, t_0; | \psi\rangle)
&=& \|~\!|\psi_\bot(t_i,t_0)\rangle\|^2
- \| (1\!\!1- |a\rangle\langle a|) U_{\Delta t}
|\psi_\bot(t_i,t_0)\rangle\|^2
\nonumber \\
&=& \langle a|U_{\Delta t}|\psi_\bot(t_i,t_0)\rangle
\langle\psi_\bot(t_i,t_0)|U^*_{\Delta t}|a\rangle
\end{eqnarray}
Using $|\psi_\bot(t_i,t_0)\rangle \langle\psi_\bot(t_i,t_0)|
\le 1\!\!1- |a\rangle \langle a|$ one obtains
\begin{eqnarray} \label{7a}
P_\bot (t_i, t_0; | \psi\rangle)
- P_\bot(t_{i+1}, t_0; | \psi\rangle)
&\le& 1 -|\langle a|U_{\Delta t}|a\rangle|^2 \nonumber \\
&=& \Delta t^2 [\langle a | HH|a\rangle
- \langle a|H|a\rangle^2]/\hbar^2+ o(\Delta t^2)
\end{eqnarray}
by Eq.~(\ref{4}). Now one can estimate, with
$t = n \Delta t + t_0, t_i = i \Delta t + t_0,$
\begin{eqnarray} \label{9}
\left| P_\bot(t,t_0; |\psi\rangle) - \|I\!\!P_\bot |\psi \rangle
\|^2 \right|
&\le& \sum_{i=1}^{n-1}~ \left| P_\bot (t_{i+1},t_0; |\psi\rangle)
- P_\bot(t_i, t_0; |\psi\rangle) \right| \nonumber\\
& & +\left| P_\bot (t_1, t_0; |\psi\rangle) -
\|I\!\!P_\bot |\psi\rangle \|^2 \right|~.
\end{eqnarray}
The sum is bounded by $(n-1) \Delta t^2 \cdot {\rm const}+
(n-1)\cdot o(\Delta t^2)$, and for
$\Delta t \rightarrow 0$ this vanishes, as does the last term on the
r.h.s. For $H = H(t)$ time-dependent, the same argument goes through
with minor modifications.

For $|a\rangle$ in the
domain of $H$ and initial state $|\psi\rangle$, this simple
argument shows that for rapidly repeated ideal measurement of
$I\!\!P_a = |a\rangle\langle a|$ the results freeze,
for $\Delta t \rightarrow 0$,
to $|a\rangle$ with probability $|\langle a|\psi\rangle |^2$ and to
$I\!\!P_\bot |\psi\rangle$ with the complementary
probability. In particular, if $| \psi\rangle = |a\rangle$, one
stays in $|a\rangle$ for $\Delta t \rightarrow 0$.

{\em Mean length of periods}. For a single system one has as results
of the measurement alternating random sequences of $a$'s and $\bot$'s
($\equiv {\rm not}~a$) of the form
\begin{equation} \label{10}
...\bot a a ...a \bot \bot .. \bot a ...
\end{equation}
The length of an $a$ sequence is defined as $\Delta t~\times$ number
of $a$'s. Similarly for $\bot$. We assume that $|a\rangle$ is not an
eigenvector of $H$, since otherwise all measurements would give
the same result, either all $a$ or all not $a$ ($\bot$).
The initial state for an $a$ sequence is $|a\rangle$ and
for an $\bot$ sequence it is
\begin{equation} \label{11}
|\phi_\bot\rangle \equiv I\!\!P_\bot U(\Delta t, 0) |a\rangle / \| \cdot \|
\end{equation}
except at the beginning when it is $|\psi\rangle$.

Starting with an $a$ the probability to have exactly $n~ a$'s
in a row, $n
\ge 1$, but not more, is by Eq.~(\ref{1}) (with $t_0 = 0$)
\begin{eqnarray} \label{12}
p_{a;n} & = & \|I\!\!P_\bot U(\Delta t,0) \psi_a (t_{n-1},0;
|a\rangle)\|^2\nonumber\\
&=& P_a(t_{n-1},0;|a\rangle ) - P_a(t_n,0;|a\rangle )
\end{eqnarray}
and analogously
\begin{equation}\label{13}
p_{\bot;n} = P_\bot (t_{n-1},0; |\phi_\bot\rangle) -
P_\bot(t_n,0;|\phi_\bot\rangle)~.
\end{equation}
The mean duration $T_a$ and $T_\bot$ of these sequences for a single
system is then, in obvious notation,
\begin{eqnarray}\label{14}
T_{a,\bot} & = & \sum^\infty_{n=1}~n \Delta t [P_{a,\bot}(t_{n-1})
-P_{a,\bot}(t_n)] \nonumber \\
&=& \sum^\infty_{n=0} \Delta t P_{a,\bot}(t_n)~.
\end{eqnarray}
 From Eq.~(\ref{2}) one obtains the exact result
\begin{eqnarray} \label{15}
T_a &=& \Delta t~\sum^\infty_{n=0}~|\langle a| U
(\Delta t,0)|a\rangle|^{2n} \nonumber \\
&=& \frac{\Delta t}{1 - |\langle a|U(\Delta t,0)|a\rangle |^2}~.
\end{eqnarray}
With Eq.~(\ref{4}) one obtains
\begin{eqnarray} \label{16}
T_a &=& \frac{1}{\Delta t} \left\{\frac{\hbar^2}{\langle a| H^2|a
\rangle -\langle a|H|a\rangle^2} + o(\Delta t^2)/\Delta t^2\right\}~.
\end{eqnarray}
The second term in the brackets becomes negligible for small $\Delta
t$, and $T_a$ diverges for $\Delta t \rightarrow 0$.
 If $|a\rangle$ is in the domain of $H^2$ then one can
replace $o(\Delta t^n)$ by $O(\Delta t^{n+1})$ where the latter
denotes terms of order at least $\Delta t^{n+1}$.

To obtain an explicit expression for $T_\bot$ we assume for simplicity
that the Hilbert space is finite-dimensional (or that $H$ is
bounded). Then one has
\begin{eqnarray} \label{17}
I\!\!P_\bot U(\Delta t,0) I\!\!P_\bot
&=& I\!\!P_\bot \, [1\!\!1-{\rm i} \Delta t H /\hbar
-\frac{1}{2} \Delta t^2 H^2/\hbar^2 + O(\Delta t^3)]\, I\!\!P_{\bot}
\nonumber\\
&=& I\!\!P_\bot {\rm e}^{-{\rm i} \Delta t I\!\!P_\bot H I\!\!P_\bot
/\hbar - \frac{1}{2} \Delta t^2 [I\!\!P_\bot H^2 I\!\!P_\bot
-(I\!\!P_\bot H I\!\!P_\bot)^2]/\hbar^2 } I\!\!P_\bot \nonumber \\
& & \times (1 + O(\Delta t^3))~.
\end{eqnarray}
Then, by Eq.~(\ref{3})
\begin{eqnarray} \label{18}
P_\bot (t_n, 0;|\psi_\bot\rangle )
&=& \langle \psi_\bot|I\!\!P_\bot
{\rm e}^{-n \Delta t^2 [I\!\!P_\bot H^2 I\!\!P_\bot
-(I\!\!P_\bot H I\!\!P_\bot)^2]/\hbar^2}
I\!\!P_\bot |\psi_\bot\rangle \nonumber \\
& & \times (1 + O(\Delta t^3))~.
\end{eqnarray}
 From this and from Eq.~(\ref{14}) one now obtains
\begin{eqnarray} \label{19}
T_\bot &=& \frac{1}{\Delta t}~\langle \phi_\bot|
\frac{\hbar^2}{I\!\!P_\bot H^2 I\!\!P_\bot - (I\!\!P_\bot H
I\!\!P_\bot)^2}|\phi_\bot \rangle
+ O(\Delta t)~. \nonumber \\
\end{eqnarray}
We note that if $|a\rangle$ is an eigenvector of $H$ then the denominators
in Eqs.~(\ref{15}) and (\ref{19}) vanish.

{\em Example}. We consider a single system with two stable levels 1 and
2. The system is driven in resonance by a classical electromagnetic
wave, e.g.~in the radio-frequency (rf) range. In the interaction
picture and with the usual rotating-wave approximation the Hamiltonian
is given by
\begin{equation} \label{20}
H ~=~ \frac{\hbar}{2}\Omega_2 \{|1\rangle\langle 2|+|2\rangle\langle 1|\}
\end{equation}
where $\Omega_2$, the so-called Rabi frequency, is proportional to the
amplitude of the driving field \cite{Loudon,Meystre}. The
time-development operator is easily calculated as
\begin{eqnarray} \label{21}
U(t,t_0) &=& \cos\frac{1}{2}\Omega_2(t - t_0)
-{\rm i}\sin\frac{1}{2}\Omega_2 (t - t_0)
\{ |1\rangle\langle 2| + |2 \rangle\langle 1|\}~.
\end{eqnarray}
 From this one finds the transition probabilities
\begin{equation} \label{22}
|\langle 2| U(t,0)|1\rangle |^2 = |\langle 1| U(t,0) |2\rangle|^2 =
\sin^2\frac{1}{2}\Omega_2 t~.
\end{equation}
For small $t$ this is quadratic in $t$. If one now determines by
repeated ideal measurements, at times $\Delta t$ apart, whether one
finds the system in state $|1\rangle$ or $|2\rangle$ one obtains a
random sequence of the form
\begin{equation} \label{23}
... 2 1 ...1 2 ...2 1 ...
\end{equation}
similar to (\ref{10}). The mean duration $T_1$ and $T_2$ of the
subsequences of $1$'s and $2$'s is given by Eq.~(\ref{15}) with $|a\rangle$
replaced by $|1\rangle$ and $|2\rangle$, respectively, and one obtains with
Eq.~(\ref{21})
\begin{equation} \label{24}
T_1 = T_2 = \frac{\Delta t}{\sin^2 \frac{1}{2} \Omega_2 \Delta t}
=\frac{4}{\Omega_2^2 \Delta t}+O(\Delta t)~.
\end{equation}
Note that $T_1 = T_2$ holds quite generally for a two-level
system, as easily seen from Eq.~(\ref{15}).
\vspace*{0.5cm}

\noindent {\bf 3. Realistic case: Light and dark periods}

\vspace*{0.5cm}

We now consider a single three-level $V$ system as in Fig.~1 and
assume the $1-2$ transition to be driven in resonance by classical
electromagnetic (rf) radiation with Rabi frequency $\Omega_2$ and
Hamiltonian as in Eq.~(\ref{20}).

We suppose that repeated measurements of level 1 are performed.
Following Refs.~\cite{Cook,Wine} we assume that each
measurement consists of a short laser (probe) pulse driving the 1-3
transition. When resonance fluorescence occurs then after the last
photon emission at the end of a probe pulse the system is in $|1\rangle$,
and when no resonance fluorescence occurs then the system was taken
by Refs.~\cite{Cook,Wine} to be in $|2\rangle$.

Experimentally one will then expect the following striking
phenomenon. One will see periods of fluorescence bursts alternating
with dark periods, as in Fig.~2. The mean duration of these light and
dark periods should be given by $T_{1,2}$ of Eq.~(\ref{24}), at least
approximately,
\begin{equation} \label{24a}
T_L \cong \frac{4}{\Omega_2^2 \Delta t}~,~T_D
\cong\frac{4}{\Omega_2^2 \Delta t}~.
\end{equation}
These periods should become longer and longer with decreasing time
$\Delta t$ between the probe pulses.

In how far the above probe pulses do indeed lead to measurements of
levels 1 and 2 and to state reduction has recently been discussed by
us in Refs.~\cite{BeHe1,BeHeSo,BeHe2} by means of the quantum jump
approach \cite{QJ}. As regards reduction, it was shown that at the
end of a probe pulse and a short transitory time the state of the
system is given either by a
density matrix extremely close, but not identical to $|1\rangle
\langle 1|$ if the
system has emitted photons, or by a density matrix very close to
$|2\rangle\langle 2|$ if no photons were emitted. After the last
photon emission
during a probe pulse the system is indeed in its ground state, but
then it may acquire a small $|2\rangle$ component until the end of the probe
pulse; its $|3\rangle$ component will decay during a short transitory
time after the pulse. When no photons are emitted the finite duration
of the probe pulse is responsible for a small $|1\rangle$
component. Hence there will be small deviations from ideal
measurements, which will lead to small corrections to the above
results.

For a probe pulse to constitute an effective measurement its duration
$\Delta \tau_{\rm p}$ has to satisfy \cite{BeHe1}
\begin{equation}\label{25}
\Delta \tau_{\rm p} \gg \max \{ A_3^{-1}, A_3 / \Omega_3^2 \}~.
\end{equation}
In addition to this one needs
\begin{equation}\label{26}
\epsilon_{\rm p} \equiv \frac{\Omega_2 A_3}{\Omega_3^2} \ll 1~,~
\epsilon_{\rm R} \equiv \frac{\Omega_2}{\Omega_3}\ll 1~,~
\epsilon_{\rm A} \equiv \frac{\Omega_2}{A_3}\ll 1~.
\end{equation}
If the time $\Delta t$ between two probe pulses satisfies
\begin{equation} \label{27}
\Delta t \gg A_3^{-1}~{\rm and}~
(\Omega_2\Delta t)^2\gg \mbox{\boldmath$\epsilon$}
\end{equation}
one can directly employ the results of Ref.~\cite{BeHeSo}. The first
of these conditions ensures that the $|3\rangle$ component has
vanished before the next pulse, the second that there are only two
possible atomic states at the end of a pulse. In case of no emission
the pulse effectively projects the system onto
\begin{eqnarray} \label{29}
\tilde{\rho}^0_{\rm P} &=& \left( \begin{array}{cc}
0 & -{\rm i} \epsilon_{\rm p} \\ {\rm i} \epsilon_{\rm p} & 1
\end{array} \right) + O(\mbox{\boldmath$\epsilon$}^2)
\end{eqnarray}
in the $|1\rangle -|2\rangle$ subspace, and in case of photon emission
onto
\begin{eqnarray} \label{30}
\tilde{\rho}_{\rm P}^> &=& \frac{1}{A_3^2 + 2\Omega_3^2
+\epsilon_{\rm p} \Omega_2 \Delta \tau_{\rm p} A_3^2}
\left( \begin{array}{cc}
A^2_3+2\Omega_3^2 & {\rm i}\epsilon_{\rm p}A^2_3
-\frac{\rm i}{2}\epsilon_{\rm A}\Omega_3^2 \\
-{\rm i}\epsilon_{\rm p}A^2_3+\frac{\rm i}{2}\epsilon_{\rm A}
\Omega_3^2 & \epsilon_{\rm p} \Omega_2 \Delta \tau_{\rm p} A_3^2
\end{array} \right)+O(\mbox{\boldmath$\epsilon$}^2)~.
\nonumber \\
\end{eqnarray}
For arbitrary initial density matrix $\rho$ the probability for no
photon emission during a probe pulse is
\begin{eqnarray} \label{28}
P_0(\Delta \tau_{\rm p};\rho) &=&
\rho_{22}-\epsilon_{\rm p} \Omega_2 \Delta \tau_{\rm p} \rho_{22}
+2\epsilon_{\rm p} \mathop{\rm Im}\rho_{12}
-2\epsilon_{\rm R} \mathop{\rm Re}\rho_{23}
+O(\mbox{\boldmath$\epsilon$}^2)~.
\end{eqnarray}

Now let $p$ be the (conditional) probability to have {\em no}
fluorescence during a pulse under the condition that there {\em had}
been fluorescence during the preceding pulse. By $q$ we denote the
probability to have {\em no} fluorescence during a pulse under the
condition that there had been {\em no} fluorescence during the
preceding pulse. In short, $p$ and $q$ are transition probabilities,
\begin{equation} \label{31}
p : {\rm yes} \rightarrow {\rm no}~~,~~q : {\rm no} \rightarrow {\rm no}~.
\end{equation}
These are the same probabilities as for the transitions from
$\tilde{\rho}^>_{\rm P}$ after a pulse to $\tilde{\rho}^0_{\rm P}$
after the next pulse and from $\tilde{\rho}^0_{\rm P}$ to
$\tilde{\rho}^0_{\rm P}$, respectively. With
\begin{equation} \label{32}
c \equiv \cos \Omega_2 \Delta t~,~ s \equiv \sin \Omega_2 \Delta t
\end{equation}
one has \cite{BeHeSo}
\begin{eqnarray} \label{33}
p &=& \frac{1}{2}(1 - c) +\epsilon_{\rm p}
\left\{2s\frac{A^2_3+\Omega_3^2}{A^2_3 + 2 \Omega_3 ^2}
+\frac{1}{2}\Omega_2\Delta \tau_{\rm p} c\frac{3 A_3^2 + 2 \Omega_3^2}
{A^2_3 + 2 \Omega^2_3}
-\frac{1}{2} \Omega_2 \Delta \tau_{\rm p}
\right\} \nonumber \\
& &-\frac{1}{2} \epsilon_{\rm A} s
\frac{\Omega_3^2}{A^2_3 + 2\Omega_3^2} + O(\mbox{\boldmath$\epsilon$}^2)~,
\\ \label{34}
q &=& \frac{1}{2}(1 + c)-\epsilon_{\rm p} \{ 2s + \frac{1}{2} \Omega_2
\Delta \tau_{\rm p} (1 + c) \} + O(\mbox{\boldmath$\epsilon$}^2)~.
\end{eqnarray}
It should be noted that for small $\Delta t$
\begin{eqnarray} \label{35}
p &=& \frac{1}{4} (\Omega_2 \Delta t)^2 + O(\mbox{\boldmath$\epsilon$})
\\ \label{36}
q &=& 1 - p + O(\mbox{\boldmath$\epsilon$})
\end{eqnarray}
and that $q \not= 1 - p$ to first order in $\mbox{\boldmath$\epsilon$}$.

The probability for a period of exactly $n$ consecutive probe pulses
{\em with} fluorescence among all such light periods is $(1 -
p)^{n-1}p$. The mean duration $T_L$ of light periods is then
\begin{equation} \label{36a}
T_L = \sum^\infty_{n=1} (\Delta \tau_{\rm p} + \Delta t)n (1-p)^{n-1} p
\end{equation}
which gives
\begin{equation}\label{37}
T_L = \frac{\Delta \tau_{\rm p} + \Delta t}{p}~.
\end{equation}
Similarly one finds for the dark periods
\begin{equation} \label{38}
T_D = \frac{\Delta \tau_{\rm p} + \Delta t}{1 - q}~.
\end{equation}
Since $1 - q$ is close, but not equal, to $p$ one has $T_L \approx
T_D$ but no longer equality. For the parameters of Ref.~\cite{Wine}
the difference is very small.

Inserting the approximate values of $p$ and $q$ from Eqs.~(\ref{35})
and (\ref{36}) one obtains
\begin{equation}\label{39}
T_L \approx T_D \approx~\frac{\Delta \tau_{\rm p} + \Delta t}
{\Delta t}~\frac{4}{\Omega_2^2 \Delta t}~.
\end{equation}
If the duration $\Delta \tau_{\rm p}$ of the probe pulse is much smaller than
the time $\Delta t$ between the pulses this agrees extremely well with
the result for ideal measurements obtained by the projection postulate
in Eqs.~(\ref{24}) and (\ref{24a}) above.

It is not possible to take the limit $\Delta t \rightarrow 0$ in
Eq.~(\ref{39}) since for the above derivation to be valid $\Delta t$
has to satisfy $\Delta t \gg A_3^{-1}$. This limit will be studied in
the next section, and we will show that $T_L$ and $T_D$ do not grow
indefinitely.
\vspace*{0.5cm}

\noindent {\bf 4. The limit of vanishing distance between probe pulses:
$\Delta \lowercase{t} \to 0$}

\vspace*{0.5cm}

To perform the limit $\Delta t \rightarrow 0$ some extra steps are
needed. For small $\Delta t$ the
population of level 3 does not vanish completely before the beginning
of the next probe pulse. Therefore, in case of fluorescence, one has
no longer a good reduction to $|1\rangle\langle 1|$ and the pulse
cannot be regarded as affecting a measurement of levels 1 and 2. In
this case the treatment of the last section has to be made more
precise by incorporating the possibly only partial decay of level 3.

Right at the end of a probe pulse -- without transient decay time --
the system is, as shown in Ref.~\cite{BeHeSo}, either in
\begin{eqnarray} \label{40}
\tilde{\rho}^0 &=& \left( \begin{array}{ccc}
0 & - {\rm i} \epsilon_{\rm p} & 0 \\
{\rm i} \epsilon_{\rm p} & 1 & - \epsilon_{\rm R} \\
0 & -\epsilon_{\rm R} & 0 \end{array} \right)
+ O(\mbox{\boldmath$\epsilon$}^2)
\end{eqnarray}
in case of no photon emission, or in
\begin{eqnarray} \label{41}
\tilde{\rho}^> &=& \frac{1}{A_3^2 + 2 \Omega^2_3 + \epsilon_{\rm p}
A^2_3 \Omega_2 \Delta \tau_{\rm p}}
\left( \begin{array}{ccc}
A^2_3 + \Omega_3^2 & {\rm i} \epsilon_{\rm p} A^2_3 & {\rm i} A_3 \Omega_3 \\
-{\rm i} \epsilon_{\rm p} A^2_3 & \epsilon_{\rm p} A^2_3 \Omega_2
\Delta \tau_{\rm p} & \epsilon_{\rm R}(A^2_3 + \Omega_3^2) \\
-{\rm i} A_3 \Omega_3 & \epsilon_{\rm R} (A^2_3 + \Omega_3^2) &
\Omega_3^2 \end{array} \right) + O(\mbox{\boldmath$\epsilon$}^2)
\nonumber \\
\end{eqnarray}
in case of fluorescence, except possibly for the {\em first} pulse of a
light period.
If the second condition in Eq. (\ref{27}) is not satisfied
by $\Delta t$ then the state at the beginning
of the first pulse in a light period
is very close to $\rho^0 $, and therefore the state $\tilde{\rho}^>$ after
the first pulse has to be calculated with initial state of the form
$\rho^0+O(\mbox{\boldmath$\epsilon$})$.
For such a state, however,
one has $1-P_0=O(\mbox{\boldmath $\epsilon$})$, by Eq. (\ref{28}),
and then $O(\mbox{\boldmath$\epsilon$}^2)$ is replaced by
$O(\mbox{\boldmath$\epsilon$})$ in Eq.~(\ref{41})
for small $\Delta t$. Thus, if the second
condition in Eq.~(\ref{27}) does not hold the first pulse in a light
period has, in principle, to be treated differently from the rest.

The transition probabilities from Eq.~(\ref{31}) are now denoted by
$\tilde{p}$ and $\tilde{q}$ and are given by
\begin{eqnarray} \label{42}
\tilde{p} &=& p-2 \epsilon_{\rm R} s\frac{\Omega_3 A_3}{A^2_3 + 2
\Omega_3^2}~{\rm e}^{-\frac{1}{2} A_3 \Delta t}
+O(\mbox{\boldmath$\epsilon$}^2) \\ \label{43}
\tilde{q} &=& q+O(\mbox{\boldmath$\epsilon$}^2)
\end{eqnarray}
with $p$ and $q$ as in Eqs.~(\ref{33}) and (\ref{34}) and $\Delta t$
arbitrary. However, for the first pulse in a light period $\tilde{p}$ is
replaced by $\tilde{p} +O(\mbox{\boldmath$\epsilon$})$.
One sees that, for $\Delta t \gg A_3^{-1}$, $ \tilde{p}$ goes
over into $p$. Eq.~(\ref{36a}) is replaced by
\begin{eqnarray}\label{43a}
T_L &=& (\Delta \tau_{\rm p} + \Delta t)(\tilde{p}+
O(\mbox{\boldmath$\epsilon$}))
+\sum^\infty_{n=2} (\Delta \tau_{\rm p} + \Delta t)n
(1-\tilde{p}+O(\mbox{\boldmath$\epsilon$}))
(1-\tilde{p})^{n-2} \tilde{p} \,\,\,
\end{eqnarray}
which gives
\begin{equation} \label{44}
T_L = \frac{\Delta \tau_{\rm p} + \Delta t}{\tilde{p}}
\end{equation}
up to terms of relative order $\mbox{\boldmath $\epsilon$}$. For
$T_D$ one obtains now
\begin{equation}\label{45}
T_D = \frac{\Delta \tau_{\rm p} + \Delta t}{1-\tilde{q}}~.
\end{equation}

Now one performs the limit $\Delta t \to 0$ and obtains
\begin{eqnarray} \label{46}
\lim_{\Delta t \rightarrow 0} \tilde{p}
&=& \epsilon_{\rm p} \Omega_2 \Delta \tau_{\rm p}
\frac{A^2_3}{A^2_3 + 2 \Omega_3^2}+O(\mbox{\boldmath$\epsilon$}^2) \nonumber\\
\lim_{\Delta t \rightarrow 0} \tilde{q}
&=& 1-\epsilon_{\rm p} \Omega_2 \Delta \tau_{\rm p}
+O(\mbox{\boldmath$\epsilon$}^2)~.
\end{eqnarray}
Inserting this into the expressions for $T_L$ and $T_D$ gives, with
$\epsilon_{\rm p} = \Omega_2 A_3/\Omega_3^2$,
\begin{eqnarray} \label{47}
\lim_{\Delta t \rightarrow 0} T_L
&=& \frac{A_3^2 + 2 \Omega^2_3}{\Omega^2_2 A^3_3}~\Omega^2_3 \nonumber \\
\lim_{\Delta t \rightarrow 0} T_D
&=& \frac{\Omega_3^2}{\Omega_2^2 A_3} ~,
\end{eqnarray}
up to terms of relative
order $\mbox{\boldmath$\epsilon$}/\Omega_2 \Delta \tau_{\rm p}$.

First of all, the limits are finite, as physically expected.
Furthermore, in the limit $\Delta t \to 0$ both driving fields are
continuously on and in this case the existence of macroscopic light
and dark periods is well known under the name `electron shelving'
\cite{Dehmelt}. The mean duration of these periods has been calculated
\cite{CT} and the result is the same as in Eq.~(\ref{47}). Thus the
continuously driven case is recovered in the limit $\Delta t \to 0$.
\vspace*{0.5cm}

\noindent {\bf 5. Conclusion}

\vspace*{0.5cm}

When applied to an ensemble of systems the QZE predicts a slow-down in
the time-development of the density matrix $\rho(t)$ under repeated
ideal measurements. An experiment to test this was performed by Itano
et al.~\cite{Wine} in which repeated state measurements were carried
out on a system with two stable levels $|1\rangle$ and $|2 \rangle$.
The measurements were implemented by short laser pulses
driving the transition from the ground state $|1\rangle$ to an
auxiliary rapidly decaying level $|3\rangle$. Occurrence or absence
of fluorescence means a system is in $|1\rangle$ or $|2\rangle$,
respectively. The experimental results indeed showed a slow-down of
the time-development of $\rho(t)$ in good agreement with the
QZE. Subsequently it was pointed out \cite{crit1} that this behavior
could be understood without recourse to any measurement theory. Indeed,
one can simply consider the probe laser as part of the dynamics and
incorporate it in the Hamiltonian or in the Bloch equations for
$\rho(t)$, never speaking of measurements. Using the quantum jump
approach \cite{QJ} (or quantum trajectories) it is possible to
understand why the dynamics is so well described by notion of
measurements and by the projection postulate \cite{BeHe1,BeHeSo}.

Instead of an ensemble of atoms we have considered
a {\em single} three-level
V system, with the same weak field driving the $|1\rangle - |2\rangle$
transition and laser pulses driving the $|1\rangle - |3\rangle$
transition as before. Taking the measurement point of view, the
projection postulate gives a quick and intuitive understanding what to
expect, namely a stochastic sequence of fluorescence bursts (light
periods) and dark periods, as in Fig.~2. Their durations should
increase with decreasing distance between the laser pulses.

Taking the dynamical point of view, Bloch equations are not so
convenient, but the quantum jump approach is particularly well adapted
to single systems. Using this approach we have shown in this paper
why, and for which parameter values, the simple projection postulate
prescription gives so highly accurate results. We have not only
calculated corrections to the projection-postulate result, but we have
also shown that if the time $\Delta t$ between the laser pulses becomes
too short then the projection postulate can no longer be applied. The
quantum jump approach, however, can also handle the limit $\Delta t
\rightarrow 0$ and yields convergence to
the well known light and dark periods of the
continuously driven system \cite{Dehmelt,CT}. These dark periods are
also called electron shelving since during this time the system is
predominantly in $|2\rangle$.
For an ensemble of many atoms different light and dark
periods will overlap, and as a result only a lower intensity of
fluorescence will be seen.

If the duration of a probe pulse becomes too short the measurement picture
is also not applicable, but the quantum jump approach still is. In
this case a numerical simulation is easiest.

In summary, we have demonstrated the usefulness of the projection
postulate for the stochastic behavior of a single system. Our dynamical
analysis also clearly shows that the projection postulate is an
idealization, sometimes even an over-idealization, and that in a more
precise treatment corrections arise. Experimentally, it should be
possible to check our results for a single ion or atom in a trap.


\newpage

\unitlength 0.6cm
\begin{picture}(18,11)
\thicklines
\put(6,9.5) {\line(1,0){3}}
\put(10,3) {\line(1,0){3}}
\put(7.5,1.5) {\line(1,0){4}}
\thinlines
\put(4.5,3.5) {\line(1,0){0.5}}
\put(5.5,3.5) {\line(1,0){0.5}}
\put(5,5.5) {\line(1,0){0.5}}
\put(5,3.5) {\line(0,1){2}}
\put(5.5,3.5) {\line(0,1){2}}
\put(7,9.5) {\vector(1,-4){2}}
\put(7.5,9.5) {\vector(1,-4){2}}
\put(9,1.5) {\vector(-1,4){2}}
\put(10,1.5) {\vector(1,1){1.5}}
\put(11.5,3) {\vector(-1,-1){1.5}}
\put (5.4,9.5){3}
\put (6,4.25){$\Longrightarrow $}
\put (6.75,6){$\Omega_3$}
\put (8.7,6){$A_3$}
\put (4,2.75){laser pulse}
\put (13.25,3){2}
\put (12.75,2){$\Longleftarrow $}
\put (14.25,2){rf field, $\Omega_2$}
\put (11.75,1.5){1}
\end{picture}

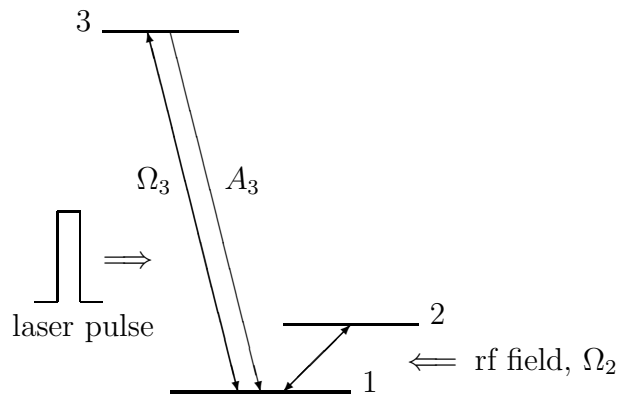
Fig. 1. V system with (meta-) stable level 2 and Einstein
coefficient $A_3$ for level 3. $\Omega_2$ and $\Omega_3$ are the Rabi
frequencies of the rf field and the probe laser, respectively.

\vspace*{1cm}

\setlength{\unitlength}{0.240900pt}
\ifx\plotpoint\undefined\newsavebox{\plotpoint}\fi
\sbox{\plotpoint}{\rule[-0.200pt]{0.400pt}{0.400pt}}%
\hspace*{0.8cm} \begin{picture}(1949,270)(200,0)
\font\gnuplot=cmr10 at 10pt
\gnuplot
\sbox{\plotpoint}{\rule[-0.200pt]{0.400pt}{0.400pt}}%
\put(176.0,113.0){\rule[-0.200pt]{411.698pt}{0.400pt}}
\put(176.0,113.0){\rule[-0.200pt]{0.400pt}{32.281pt}}
\put(176.0,113.0){\rule[-0.200pt]{0.400pt}{4.818pt}}
\put(176,68){\makebox(0,0){0}}
\put(176.0,227.0){\rule[-0.200pt]{0.400pt}{4.818pt}}
\put(390.0,113.0){\rule[-0.200pt]{0.400pt}{4.818pt}}
\put(390,68){\makebox(0,0){5}}
\put(390.0,227.0){\rule[-0.200pt]{0.400pt}{4.818pt}}
\put(603.0,113.0){\rule[-0.200pt]{0.400pt}{4.818pt}}
\put(603,68){\makebox(0,0){10}}
\put(603.0,227.0){\rule[-0.200pt]{0.400pt}{4.818pt}}
\put(817.0,113.0){\rule[-0.200pt]{0.400pt}{4.818pt}}
\put(817,68){\makebox(0,0){15}}
\put(817.0,227.0){\rule[-0.200pt]{0.400pt}{4.818pt}}
\put(1031.0,113.0){\rule[-0.200pt]{0.400pt}{4.818pt}}
\put(1031,68){\makebox(0,0){20}}
\put(1031.0,227.0){\rule[-0.200pt]{0.400pt}{4.818pt}}
\put(1244.0,113.0){\rule[-0.200pt]{0.400pt}{4.818pt}}
\put(1244,68){\makebox(0,0){25}}
\put(1244.0,227.0){\rule[-0.200pt]{0.400pt}{4.818pt}}
\put(1458.0,113.0){\rule[-0.200pt]{0.400pt}{4.818pt}}
\put(1458,68){\makebox(0,0){30}}
\put(1458.0,227.0){\rule[-0.200pt]{0.400pt}{4.818pt}}
\put(1671.0,113.0){\rule[-0.200pt]{0.400pt}{4.818pt}}
\put(1671,68){\makebox(0,0){35}}
\put(1671.0,227.0){\rule[-0.200pt]{0.400pt}{4.818pt}}
\put(1885.0,113.0){\rule[-0.200pt]{0.400pt}{4.818pt}}
\put(1885,68){\makebox(0,0){40}}
\put(1885.0,227.0){\rule[-0.200pt]{0.400pt}{4.818pt}}
\put(176.0,113.0){\rule[-0.200pt]{411.698pt}{0.400pt}}
\put(1885.0,113.0){\rule[-0.200pt]{0.400pt}{32.281pt}}
\put(176.0,247.0){\rule[-0.200pt]{411.698pt}{0.400pt}}
\put(1030,0){\makebox(0,0){(a)
$ \Delta t = T_\pi/2 $\hspace{5cm} $ t ~[ T_{\rm \pi } ] $}}
\put(176.0,113.0){\rule[-0.200pt]{0.400pt}{32.281pt}}
\put(197.0,113.0){\rule[-0.200pt]{0.400pt}{16.140pt}}
\put(219,113){\usebox{\plotpoint}}
\put(240,113){\usebox{\plotpoint}}
\put(261,113){\usebox{\plotpoint}}
\put(283.0,113.0){\rule[-0.200pt]{0.400pt}{16.140pt}}
\put(304.0,113.0){\rule[-0.200pt]{0.400pt}{16.140pt}}
\put(326.0,113.0){\rule[-0.200pt]{0.400pt}{16.140pt}}
\put(347.0,113.0){\rule[-0.200pt]{0.400pt}{16.140pt}}
\put(368,113){\usebox{\plotpoint}}
\put(390,113){\usebox{\plotpoint}}
\put(411.0,113.0){\rule[-0.200pt]{0.400pt}{16.140pt}}
\put(432,113){\usebox{\plotpoint}}
\put(454.0,113.0){\rule[-0.200pt]{0.400pt}{16.140pt}}
\put(475.0,113.0){\rule[-0.200pt]{0.400pt}{16.140pt}}
\put(496.0,113.0){\rule[-0.200pt]{0.400pt}{16.140pt}}
\put(518,113){\usebox{\plotpoint}}
\put(539,113){\usebox{\plotpoint}}
\put(561.0,113.0){\rule[-0.200pt]{0.400pt}{16.140pt}}
\put(582.0,113.0){\rule[-0.200pt]{0.400pt}{16.140pt}}
\put(603.0,113.0){\rule[-0.200pt]{0.400pt}{16.140pt}}
\put(625.0,113.0){\rule[-0.200pt]{0.400pt}{16.140pt}}
\put(646.0,113.0){\rule[-0.200pt]{0.400pt}{16.140pt}}
\put(667,113){\usebox{\plotpoint}}
\put(689,113){\usebox{\plotpoint}}
\put(710.0,113.0){\rule[-0.200pt]{0.400pt}{16.140pt}}
\put(731,113){\usebox{\plotpoint}}
\put(753.0,113.0){\rule[-0.200pt]{0.400pt}{16.140pt}}
\put(774,113){\usebox{\plotpoint}}
\put(796.0,113.0){\rule[-0.200pt]{0.400pt}{16.140pt}}
\put(817.0,113.0){\rule[-0.200pt]{0.400pt}{16.140pt}}
\put(838.0,113.0){\rule[-0.200pt]{0.400pt}{16.140pt}}
\put(860,113){\usebox{\plotpoint}}
\put(881.0,113.0){\rule[-0.200pt]{0.400pt}{16.140pt}}
\put(902.0,113.0){\rule[-0.200pt]{0.400pt}{16.140pt}}
\put(924,113){\usebox{\plotpoint}}
\put(945.0,113.0){\rule[-0.200pt]{0.400pt}{16.140pt}}
\put(966.0,113.0){\rule[-0.200pt]{0.400pt}{16.140pt}}
\put(988,113){\usebox{\plotpoint}}
\put(1009,113){\usebox{\plotpoint}}
\put(1031.0,113.0){\rule[-0.200pt]{0.400pt}{16.140pt}}
\put(1052,113){\usebox{\plotpoint}}
\put(1073,113){\usebox{\plotpoint}}
\put(1095.0,113.0){\rule[-0.200pt]{0.400pt}{16.140pt}}
\put(1116.0,113.0){\rule[-0.200pt]{0.400pt}{16.140pt}}
\put(1137.0,113.0){\rule[-0.200pt]{0.400pt}{16.140pt}}
\put(1159.0,113.0){\rule[-0.200pt]{0.400pt}{16.140pt}}
\put(1180,113){\usebox{\plotpoint}}
\put(1201,113){\usebox{\plotpoint}}
\put(1223.0,113.0){\rule[-0.200pt]{0.400pt}{16.140pt}}
\put(1244.0,113.0){\rule[-0.200pt]{0.400pt}{16.140pt}}
\put(1265.0,113.0){\rule[-0.200pt]{0.400pt}{16.140pt}}
\put(1287.0,113.0){\rule[-0.200pt]{0.400pt}{16.140pt}}
\put(1308.0,113.0){\rule[-0.200pt]{0.400pt}{16.140pt}}
\put(1330.0,113.0){\rule[-0.200pt]{0.400pt}{16.140pt}}
\put(1351,113){\usebox{\plotpoint}}
\put(1372,113){\usebox{\plotpoint}}
\put(1394,113){\usebox{\plotpoint}}
\put(1415.0,113.0){\rule[-0.200pt]{0.400pt}{16.140pt}}
\put(1436.0,113.0){\rule[-0.200pt]{0.400pt}{16.140pt}}
\put(1458,113){\usebox{\plotpoint}}
\put(1479,113){\usebox{\plotpoint}}
\put(1500,113){\usebox{\plotpoint}}
\put(1522.0,113.0){\rule[-0.200pt]{0.400pt}{16.140pt}}
\put(1543,113){\usebox{\plotpoint}}
\put(1565,113){\usebox{\plotpoint}}
\put(1586.0,113.0){\rule[-0.200pt]{0.400pt}{16.140pt}}
\put(1607.0,113.0){\rule[-0.200pt]{0.400pt}{16.140pt}}
\put(1629,113){\usebox{\plotpoint}}
\put(1650.0,113.0){\rule[-0.200pt]{0.400pt}{16.140pt}}
\put(1671.0,113.0){\rule[-0.200pt]{0.400pt}{16.140pt}}
\put(1693,113){\usebox{\plotpoint}}
\put(1714.0,113.0){\rule[-0.200pt]{0.400pt}{16.140pt}}
\put(1735.0,113.0){\rule[-0.200pt]{0.400pt}{16.140pt}}
\put(1757.0,113.0){\rule[-0.200pt]{0.400pt}{16.140pt}}
\put(1778.0,113.0){\rule[-0.200pt]{0.400pt}{16.140pt}}
\put(1800.0,113.0){\rule[-0.200pt]{0.400pt}{16.140pt}}
\put(1821,113){\usebox{\plotpoint}}
\put(1842.0,113.0){\rule[-0.200pt]{0.400pt}{16.140pt}}
\put(1864.0,113.0){\rule[-0.200pt]{0.400pt}{16.140pt}}
\put(1885.0,113.0){\rule[-0.200pt]{0.400pt}{16.140pt}}
\end{picture}

\vspace*{0.4cm}
\setlength{\unitlength}{0.240900pt}
\ifx\plotpoint\undefined\newsavebox{\plotpoint}\fi
\sbox{\plotpoint}{\rule[-0.200pt]{0.400pt}{0.400pt}}%
\hspace*{0.8cm} \begin{picture}(1949,270)(200,0)
\font\gnuplot=cmr10 at 10pt
\gnuplot
\sbox{\plotpoint}{\rule[-0.200pt]{0.400pt}{0.400pt}}%
\put(176.0,113.0){\rule[-0.200pt]{411.698pt}{0.400pt}}
\put(176.0,113.0){\rule[-0.200pt]{0.400pt}{32.281pt}}
\put(176.0,113.0){\rule[-0.200pt]{0.400pt}{4.818pt}}
\put(176,68){\makebox(0,0){0}}
\put(176.0,227.0){\rule[-0.200pt]{0.400pt}{4.818pt}}
\put(390.0,113.0){\rule[-0.200pt]{0.400pt}{4.818pt}}
\put(390,68){\makebox(0,0){5}}
\put(390.0,227.0){\rule[-0.200pt]{0.400pt}{4.818pt}}
\put(603.0,113.0){\rule[-0.200pt]{0.400pt}{4.818pt}}
\put(603,68){\makebox(0,0){10}}
\put(603.0,227.0){\rule[-0.200pt]{0.400pt}{4.818pt}}
\put(817.0,113.0){\rule[-0.200pt]{0.400pt}{4.818pt}}
\put(817,68){\makebox(0,0){15}}
\put(817.0,227.0){\rule[-0.200pt]{0.400pt}{4.818pt}}
\put(1031.0,113.0){\rule[-0.200pt]{0.400pt}{4.818pt}}
\put(1031,68){\makebox(0,0){20}}
\put(1031.0,227.0){\rule[-0.200pt]{0.400pt}{4.818pt}}
\put(1244.0,113.0){\rule[-0.200pt]{0.400pt}{4.818pt}}
\put(1244,68){\makebox(0,0){25}}
\put(1244.0,227.0){\rule[-0.200pt]{0.400pt}{4.818pt}}
\put(1458.0,113.0){\rule[-0.200pt]{0.400pt}{4.818pt}}
\put(1458,68){\makebox(0,0){30}}
\put(1458.0,227.0){\rule[-0.200pt]{0.400pt}{4.818pt}}
\put(1671.0,113.0){\rule[-0.200pt]{0.400pt}{4.818pt}}
\put(1671,68){\makebox(0,0){35}}
\put(1671.0,227.0){\rule[-0.200pt]{0.400pt}{4.818pt}}
\put(1885.0,113.0){\rule[-0.200pt]{0.400pt}{4.818pt}}
\put(1885,68){\makebox(0,0){40}}
\put(1885.0,227.0){\rule[-0.200pt]{0.400pt}{4.818pt}}
\put(176.0,113.0){\rule[-0.200pt]{411.698pt}{0.400pt}}
\put(1885.0,113.0){\rule[-0.200pt]{0.400pt}{32.281pt}}
\put(176.0,247.0){\rule[-0.200pt]{411.698pt}{0.400pt}}
\put(1030,0){\makebox(0,0){(b)
$ \Delta t = T_\pi/4 $\hspace{5cm} $ t ~[ T_{\rm \pi } ] $}}
\put(176.0,113.0){\rule[-0.200pt]{0.400pt}{32.281pt}}
\put(187.0,113.0){\rule[-0.200pt]{0.400pt}{16.140pt}}
\put(197.0,113.0){\rule[-0.200pt]{0.400pt}{16.140pt}}
\put(208.0,113.0){\rule[-0.200pt]{0.400pt}{16.140pt}}
\put(219.0,113.0){\rule[-0.200pt]{0.400pt}{16.140pt}}
\put(229.0,113.0){\rule[-0.200pt]{0.400pt}{16.140pt}}
\put(240.0,113.0){\rule[-0.200pt]{0.400pt}{16.140pt}}
\put(251.0,113.0){\rule[-0.200pt]{0.400pt}{16.140pt}}
\put(261.0,113.0){\rule[-0.200pt]{0.400pt}{16.140pt}}
\put(272.0,113.0){\rule[-0.200pt]{0.400pt}{16.140pt}}
\put(283,113){\usebox{\plotpoint}}
\put(293,113){\usebox{\plotpoint}}
\put(304,113){\usebox{\plotpoint}}
\put(315,113){\usebox{\plotpoint}}
\put(326.0,113.0){\rule[-0.200pt]{0.400pt}{16.140pt}}
\put(336.0,113.0){\rule[-0.200pt]{0.400pt}{16.140pt}}
\put(347.0,113.0){\rule[-0.200pt]{0.400pt}{16.140pt}}
\put(358.0,113.0){\rule[-0.200pt]{0.400pt}{16.140pt}}
\put(368.0,113.0){\rule[-0.200pt]{0.400pt}{16.140pt}}
\put(379.0,113.0){\rule[-0.200pt]{0.400pt}{16.140pt}}
\put(390.0,113.0){\rule[-0.200pt]{0.400pt}{16.140pt}}
\put(400.0,113.0){\rule[-0.200pt]{0.400pt}{16.140pt}}
\put(411.0,113.0){\rule[-0.200pt]{0.400pt}{16.140pt}}
\put(422.0,113.0){\rule[-0.200pt]{0.400pt}{16.140pt}}
\put(432.0,113.0){\rule[-0.200pt]{0.400pt}{16.140pt}}
\put(443.0,113.0){\rule[-0.200pt]{0.400pt}{16.140pt}}
\put(454,113){\usebox{\plotpoint}}
\put(464,113){\usebox{\plotpoint}}
\put(475,113){\usebox{\plotpoint}}
\put(486,113){\usebox{\plotpoint}}
\put(496,113){\usebox{\plotpoint}}
\put(507,113){\usebox{\plotpoint}}
\put(518,113){\usebox{\plotpoint}}
\put(528.0,113.0){\rule[-0.200pt]{0.400pt}{16.140pt}}
\put(539.0,113.0){\rule[-0.200pt]{0.400pt}{16.140pt}}
\put(550,113){\usebox{\plotpoint}}
\put(561,113){\usebox{\plotpoint}}
\put(571,113){\usebox{\plotpoint}}
\put(582,113){\usebox{\plotpoint}}
\put(593,113){\usebox{\plotpoint}}
\put(603,113){\usebox{\plotpoint}}
\put(614,113){\usebox{\plotpoint}}
\put(625,113){\usebox{\plotpoint}}
\put(635.0,113.0){\rule[-0.200pt]{0.400pt}{16.140pt}}
\put(646.0,113.0){\rule[-0.200pt]{0.400pt}{16.140pt}}
\put(657.0,113.0){\rule[-0.200pt]{0.400pt}{16.140pt}}
\put(667.0,113.0){\rule[-0.200pt]{0.400pt}{16.140pt}}
\put(678.0,113.0){\rule[-0.200pt]{0.400pt}{16.140pt}}
\put(689.0,113.0){\rule[-0.200pt]{0.400pt}{16.140pt}}
\put(699.0,113.0){\rule[-0.200pt]{0.400pt}{16.140pt}}
\put(710.0,113.0){\rule[-0.200pt]{0.400pt}{16.140pt}}
\put(721.0,113.0){\rule[-0.200pt]{0.400pt}{16.140pt}}
\put(731.0,113.0){\rule[-0.200pt]{0.400pt}{16.140pt}}
\put(742.0,113.0){\rule[-0.200pt]{0.400pt}{16.140pt}}
\put(753.0,113.0){\rule[-0.200pt]{0.400pt}{16.140pt}}
\put(763.0,113.0){\rule[-0.200pt]{0.400pt}{16.140pt}}
\put(774.0,113.0){\rule[-0.200pt]{0.400pt}{16.140pt}}
\put(785,113){\usebox{\plotpoint}}
\put(796.0,113.0){\rule[-0.200pt]{0.400pt}{16.140pt}}
\put(806.0,113.0){\rule[-0.200pt]{0.400pt}{16.140pt}}
\put(817.0,113.0){\rule[-0.200pt]{0.400pt}{16.140pt}}
\put(828.0,113.0){\rule[-0.200pt]{0.400pt}{16.140pt}}
\put(838.0,113.0){\rule[-0.200pt]{0.400pt}{16.140pt}}
\put(849.0,113.0){\rule[-0.200pt]{0.400pt}{16.140pt}}
\put(860.0,113.0){\rule[-0.200pt]{0.400pt}{16.140pt}}
\put(870.0,113.0){\rule[-0.200pt]{0.400pt}{16.140pt}}
\put(881.0,113.0){\rule[-0.200pt]{0.400pt}{16.140pt}}
\put(892.0,113.0){\rule[-0.200pt]{0.400pt}{16.140pt}}
\put(902.0,113.0){\rule[-0.200pt]{0.400pt}{16.140pt}}
\put(913.0,113.0){\rule[-0.200pt]{0.400pt}{16.140pt}}
\put(924.0,113.0){\rule[-0.200pt]{0.400pt}{16.140pt}}
\put(934.0,113.0){\rule[-0.200pt]{0.400pt}{16.140pt}}
\put(945.0,113.0){\rule[-0.200pt]{0.400pt}{16.140pt}}
\put(956.0,113.0){\rule[-0.200pt]{0.400pt}{16.140pt}}
\put(966.0,113.0){\rule[-0.200pt]{0.400pt}{16.140pt}}
\put(977.0,113.0){\rule[-0.200pt]{0.400pt}{16.140pt}}
\put(988.0,113.0){\rule[-0.200pt]{0.400pt}{16.140pt}}
\put(998.0,113.0){\rule[-0.200pt]{0.400pt}{16.140pt}}
\put(1009.0,113.0){\rule[-0.200pt]{0.400pt}{16.140pt}}
\put(1020.0,113.0){\rule[-0.200pt]{0.400pt}{16.140pt}}
\put(1031.0,113.0){\rule[-0.200pt]{0.400pt}{16.140pt}}
\put(1041.0,113.0){\rule[-0.200pt]{0.400pt}{16.140pt}}
\put(1052.0,113.0){\rule[-0.200pt]{0.400pt}{16.140pt}}
\put(1063,113){\usebox{\plotpoint}}
\put(1073,113){\usebox{\plotpoint}}
\put(1084,113){\usebox{\plotpoint}}
\put(1095.0,113.0){\rule[-0.200pt]{0.400pt}{16.140pt}}
\put(1105.0,113.0){\rule[-0.200pt]{0.400pt}{16.140pt}}
\put(1116.0,113.0){\rule[-0.200pt]{0.400pt}{16.140pt}}
\put(1127.0,113.0){\rule[-0.200pt]{0.400pt}{16.140pt}}
\put(1137.0,113.0){\rule[-0.200pt]{0.400pt}{16.140pt}}
\put(1148.0,113.0){\rule[-0.200pt]{0.400pt}{16.140pt}}
\put(1159.0,113.0){\rule[-0.200pt]{0.400pt}{16.140pt}}
\put(1169,113){\usebox{\plotpoint}}
\put(1180,113){\usebox{\plotpoint}}
\put(1191,113){\usebox{\plotpoint}}
\put(1201,113){\usebox{\plotpoint}}
\put(1212,113){\usebox{\plotpoint}}
\put(1223,113){\usebox{\plotpoint}}
\put(1233,113){\usebox{\plotpoint}}
\put(1244,113){\usebox{\plotpoint}}
\put(1255,113){\usebox{\plotpoint}}
\put(1265.0,113.0){\rule[-0.200pt]{0.400pt}{16.140pt}}
\put(1276.0,113.0){\rule[-0.200pt]{0.400pt}{16.140pt}}
\put(1287.0,113.0){\rule[-0.200pt]{0.400pt}{16.140pt}}
\put(1298.0,113.0){\rule[-0.200pt]{0.400pt}{16.140pt}}
\put(1308.0,113.0){\rule[-0.200pt]{0.400pt}{16.140pt}}
\put(1319,113){\usebox{\plotpoint}}
\put(1330,113){\usebox{\plotpoint}}
\put(1340,113){\usebox{\plotpoint}}
\put(1351,113){\usebox{\plotpoint}}
\put(1362,113){\usebox{\plotpoint}}
\put(1372,113){\usebox{\plotpoint}}
\put(1383,113){\usebox{\plotpoint}}
\put(1394,113){\usebox{\plotpoint}}
\put(1404,113){\usebox{\plotpoint}}
\put(1415,113){\usebox{\plotpoint}}
\put(1426,113){\usebox{\plotpoint}}
\put(1436,113){\usebox{\plotpoint}}
\put(1447,113){\usebox{\plotpoint}}
\put(1458,113){\usebox{\plotpoint}}
\put(1468,113){\usebox{\plotpoint}}
\put(1479.0,113.0){\rule[-0.200pt]{0.400pt}{16.140pt}}
\put(1490.0,113.0){\rule[-0.200pt]{0.400pt}{16.140pt}}
\put(1500,113){\usebox{\plotpoint}}
\put(1511,113){\usebox{\plotpoint}}
\put(1522,113){\usebox{\plotpoint}}
\put(1533,113){\usebox{\plotpoint}}
\put(1543.0,113.0){\rule[-0.200pt]{0.400pt}{16.140pt}}
\put(1554,113){\usebox{\plotpoint}}
\put(1565,113){\usebox{\plotpoint}}
\put(1575,113){\usebox{\plotpoint}}
\put(1586,113){\usebox{\plotpoint}}
\put(1597,113){\usebox{\plotpoint}}
\put(1607,113){\usebox{\plotpoint}}
\put(1618,113){\usebox{\plotpoint}}
\put(1629,113){\usebox{\plotpoint}}
\put(1639,113){\usebox{\plotpoint}}
\put(1650,113){\usebox{\plotpoint}}
\put(1661,113){\usebox{\plotpoint}}
\put(1671,113){\usebox{\plotpoint}}
\put(1682,113){\usebox{\plotpoint}}
\put(1693.0,113.0){\rule[-0.200pt]{0.400pt}{16.140pt}}
\put(1703.0,113.0){\rule[-0.200pt]{0.400pt}{16.140pt}}
\put(1714.0,113.0){\rule[-0.200pt]{0.400pt}{16.140pt}}
\put(1725.0,113.0){\rule[-0.200pt]{0.400pt}{16.140pt}}
\put(1735.0,113.0){\rule[-0.200pt]{0.400pt}{16.140pt}}
\put(1746.0,113.0){\rule[-0.200pt]{0.400pt}{16.140pt}}
\put(1757.0,113.0){\rule[-0.200pt]{0.400pt}{16.140pt}}
\put(1768,113){\usebox{\plotpoint}}
\put(1778,113){\usebox{\plotpoint}}
\put(1789,113){\usebox{\plotpoint}}
\put(1800,113){\usebox{\plotpoint}}
\put(1810,113){\usebox{\plotpoint}}
\put(1821,113){\usebox{\plotpoint}}
\put(1832,113){\usebox{\plotpoint}}
\put(1842,113){\usebox{\plotpoint}}
\put(1853.0,113.0){\rule[-0.200pt]{0.400pt}{16.140pt}}
\put(1864.0,113.0){\rule[-0.200pt]{0.400pt}{16.140pt}}
\put(1874.0,113.0){\rule[-0.200pt]{0.400pt}{16.140pt}}
\put(1885.0,113.0){\rule[-0.200pt]{0.400pt}{16.140pt}}
\end{picture}

\vspace*{0.4cm}
\setlength{\unitlength}{0.240900pt}
\ifx\plotpoint\undefined\newsavebox{\plotpoint}\fi
\sbox{\plotpoint}{\rule[-0.200pt]{0.400pt}{0.400pt}}%
\hspace*{0.8cm} \begin{picture}(1949,270)(200,0)
\font\gnuplot=cmr10 at 10pt
\gnuplot
\sbox{\plotpoint}{\rule[-0.200pt]{0.400pt}{0.400pt}}%
\put(176.0,113.0){\rule[-0.200pt]{411.698pt}{0.400pt}}
\put(176.0,113.0){\rule[-0.200pt]{0.400pt}{32.281pt}}
\put(176.0,113.0){\rule[-0.200pt]{0.400pt}{4.818pt}}
\put(176,68){\makebox(0,0){0}}
\put(176.0,227.0){\rule[-0.200pt]{0.400pt}{4.818pt}}
\put(390.0,113.0){\rule[-0.200pt]{0.400pt}{4.818pt}}
\put(390,68){\makebox(0,0){5}}
\put(390.0,227.0){\rule[-0.200pt]{0.400pt}{4.818pt}}
\put(603.0,113.0){\rule[-0.200pt]{0.400pt}{4.818pt}}
\put(603,68){\makebox(0,0){10}}
\put(603.0,227.0){\rule[-0.200pt]{0.400pt}{4.818pt}}
\put(817.0,113.0){\rule[-0.200pt]{0.400pt}{4.818pt}}
\put(817,68){\makebox(0,0){15}}
\put(817.0,227.0){\rule[-0.200pt]{0.400pt}{4.818pt}}
\put(1031.0,113.0){\rule[-0.200pt]{0.400pt}{4.818pt}}
\put(1031,68){\makebox(0,0){20}}
\put(1031.0,227.0){\rule[-0.200pt]{0.400pt}{4.818pt}}
\put(1244.0,113.0){\rule[-0.200pt]{0.400pt}{4.818pt}}
\put(1244,68){\makebox(0,0){25}}
\put(1244.0,227.0){\rule[-0.200pt]{0.400pt}{4.818pt}}
\put(1458.0,113.0){\rule[-0.200pt]{0.400pt}{4.818pt}}
\put(1458,68){\makebox(0,0){30}}
\put(1458.0,227.0){\rule[-0.200pt]{0.400pt}{4.818pt}}
\put(1671.0,113.0){\rule[-0.200pt]{0.400pt}{4.818pt}}
\put(1671,68){\makebox(0,0){35}}
\put(1671.0,227.0){\rule[-0.200pt]{0.400pt}{4.818pt}}
\put(1885.0,113.0){\rule[-0.200pt]{0.400pt}{4.818pt}}
\put(1885,68){\makebox(0,0){40}}
\put(1885.0,227.0){\rule[-0.200pt]{0.400pt}{4.818pt}}
\put(176.0,113.0){\rule[-0.200pt]{411.698pt}{0.400pt}}
\put(1885.0,113.0){\rule[-0.200pt]{0.400pt}{32.281pt}}
\put(176.0,247.0){\rule[-0.200pt]{411.698pt}{0.400pt}}
\put(1030,0){\makebox(0,0){(c)
$\Delta t = T_\pi/6 $\hspace{5cm} $ t ~[ T_{\rm \pi } ] $}}
\put(176.0,113.0){\rule[-0.200pt]{0.400pt}{32.281pt}}
\put(183.0,113.0){\rule[-0.200pt]{0.400pt}{16.140pt}}
\put(190.0,113.0){\rule[-0.200pt]{0.400pt}{16.140pt}}
\put(197.0,113.0){\rule[-0.200pt]{0.400pt}{16.140pt}}
\put(204.0,113.0){\rule[-0.200pt]{0.400pt}{16.140pt}}
\put(212.0,113.0){\rule[-0.200pt]{0.400pt}{16.140pt}}
\put(219.0,113.0){\rule[-0.200pt]{0.400pt}{16.140pt}}
\put(226.0,113.0){\rule[-0.200pt]{0.400pt}{16.140pt}}
\put(233.0,113.0){\rule[-0.200pt]{0.400pt}{16.140pt}}
\put(240.0,113.0){\rule[-0.200pt]{0.400pt}{16.140pt}}
\put(247.0,113.0){\rule[-0.200pt]{0.400pt}{16.140pt}}
\put(254.0,113.0){\rule[-0.200pt]{0.400pt}{16.140pt}}
\put(261.0,113.0){\rule[-0.200pt]{0.400pt}{16.140pt}}
\put(269.0,113.0){\rule[-0.200pt]{0.400pt}{16.140pt}}
\put(276.0,113.0){\rule[-0.200pt]{0.400pt}{16.140pt}}
\put(283.0,113.0){\rule[-0.200pt]{0.400pt}{16.140pt}}
\put(290.0,113.0){\rule[-0.200pt]{0.400pt}{16.140pt}}
\put(297.0,113.0){\rule[-0.200pt]{0.400pt}{16.140pt}}
\put(304.0,113.0){\rule[-0.200pt]{0.400pt}{16.140pt}}
\put(311.0,113.0){\rule[-0.200pt]{0.400pt}{16.140pt}}
\put(318.0,113.0){\rule[-0.200pt]{0.400pt}{16.140pt}}
\put(326.0,113.0){\rule[-0.200pt]{0.400pt}{16.140pt}}
\put(333.0,113.0){\rule[-0.200pt]{0.400pt}{16.140pt}}
\put(340.0,113.0){\rule[-0.200pt]{0.400pt}{16.140pt}}
\put(347.0,113.0){\rule[-0.200pt]{0.400pt}{16.140pt}}
\put(354.0,113.0){\rule[-0.200pt]{0.400pt}{16.140pt}}
\put(361,113){\usebox{\plotpoint}}
\put(368,113){\usebox{\plotpoint}}
\put(375,113){\usebox{\plotpoint}}
\put(383,113){\usebox{\plotpoint}}
\put(390,113){\usebox{\plotpoint}}
\put(397,113){\usebox{\plotpoint}}
\put(404,113){\usebox{\plotpoint}}
\put(411.0,113.0){\rule[-0.200pt]{0.400pt}{16.140pt}}
\put(418.0,113.0){\rule[-0.200pt]{0.400pt}{16.140pt}}
\put(425,113){\usebox{\plotpoint}}
\put(432,113){\usebox{\plotpoint}}
\put(439,113){\usebox{\plotpoint}}
\put(447,113){\usebox{\plotpoint}}
\put(454,113){\usebox{\plotpoint}}
\put(461,113){\usebox{\plotpoint}}
\put(468,113){\usebox{\plotpoint}}
\put(475,113){\usebox{\plotpoint}}
\put(482.0,113.0){\rule[-0.200pt]{0.400pt}{16.140pt}}
\put(489.0,113.0){\rule[-0.200pt]{0.400pt}{16.140pt}}
\put(496.0,113.0){\rule[-0.200pt]{0.400pt}{16.140pt}}
\put(504.0,113.0){\rule[-0.200pt]{0.400pt}{16.140pt}}
\put(511.0,113.0){\rule[-0.200pt]{0.400pt}{16.140pt}}
\put(518.0,113.0){\rule[-0.200pt]{0.400pt}{16.140pt}}
\put(525.0,113.0){\rule[-0.200pt]{0.400pt}{16.140pt}}
\put(532.0,113.0){\rule[-0.200pt]{0.400pt}{16.140pt}}
\put(539.0,113.0){\rule[-0.200pt]{0.400pt}{16.140pt}}
\put(546.0,113.0){\rule[-0.200pt]{0.400pt}{16.140pt}}
\put(553.0,113.0){\rule[-0.200pt]{0.400pt}{16.140pt}}
\put(561.0,113.0){\rule[-0.200pt]{0.400pt}{16.140pt}}
\put(568.0,113.0){\rule[-0.200pt]{0.400pt}{16.140pt}}
\put(575.0,113.0){\rule[-0.200pt]{0.400pt}{16.140pt}}
\put(582,113){\usebox{\plotpoint}}
\put(589.0,113.0){\rule[-0.200pt]{0.400pt}{16.140pt}}
\put(596.0,113.0){\rule[-0.200pt]{0.400pt}{16.140pt}}
\put(603.0,113.0){\rule[-0.200pt]{0.400pt}{16.140pt}}
\put(610.0,113.0){\rule[-0.200pt]{0.400pt}{16.140pt}}
\put(617.0,113.0){\rule[-0.200pt]{0.400pt}{16.140pt}}
\put(625.0,113.0){\rule[-0.200pt]{0.400pt}{16.140pt}}
\put(632.0,113.0){\rule[-0.200pt]{0.400pt}{16.140pt}}
\put(639.0,113.0){\rule[-0.200pt]{0.400pt}{16.140pt}}
\put(646.0,113.0){\rule[-0.200pt]{0.400pt}{16.140pt}}
\put(653.0,113.0){\rule[-0.200pt]{0.400pt}{16.140pt}}
\put(660.0,113.0){\rule[-0.200pt]{0.400pt}{16.140pt}}
\put(667.0,113.0){\rule[-0.200pt]{0.400pt}{16.140pt}}
\put(674.0,113.0){\rule[-0.200pt]{0.400pt}{16.140pt}}
\put(682.0,113.0){\rule[-0.200pt]{0.400pt}{16.140pt}}
\put(689.0,113.0){\rule[-0.200pt]{0.400pt}{16.140pt}}
\put(696.0,113.0){\rule[-0.200pt]{0.400pt}{16.140pt}}
\put(703.0,113.0){\rule[-0.200pt]{0.400pt}{16.140pt}}
\put(710.0,113.0){\rule[-0.200pt]{0.400pt}{16.140pt}}
\put(717.0,113.0){\rule[-0.200pt]{0.400pt}{16.140pt}}
\put(724.0,113.0){\rule[-0.200pt]{0.400pt}{16.140pt}}
\put(731.0,113.0){\rule[-0.200pt]{0.400pt}{16.140pt}}
\put(739.0,113.0){\rule[-0.200pt]{0.400pt}{16.140pt}}
\put(746.0,113.0){\rule[-0.200pt]{0.400pt}{16.140pt}}
\put(753.0,113.0){\rule[-0.200pt]{0.400pt}{16.140pt}}
\put(760.0,113.0){\rule[-0.200pt]{0.400pt}{16.140pt}}
\put(767.0,113.0){\rule[-0.200pt]{0.400pt}{16.140pt}}
\put(774.0,113.0){\rule[-0.200pt]{0.400pt}{16.140pt}}
\put(781.0,113.0){\rule[-0.200pt]{0.400pt}{16.140pt}}
\put(788.0,113.0){\rule[-0.200pt]{0.400pt}{16.140pt}}
\put(796.0,113.0){\rule[-0.200pt]{0.400pt}{16.140pt}}
\put(803.0,113.0){\rule[-0.200pt]{0.400pt}{16.140pt}}
\put(810.0,113.0){\rule[-0.200pt]{0.400pt}{16.140pt}}
\put(817.0,113.0){\rule[-0.200pt]{0.400pt}{16.140pt}}
\put(824.0,113.0){\rule[-0.200pt]{0.400pt}{16.140pt}}
\put(831.0,113.0){\rule[-0.200pt]{0.400pt}{16.140pt}}
\put(838,113){\usebox{\plotpoint}}
\put(845,113){\usebox{\plotpoint}}
\put(852,113){\usebox{\plotpoint}}
\put(860,113){\usebox{\plotpoint}}
\put(867,113){\usebox{\plotpoint}}
\put(874,113){\usebox{\plotpoint}}
\put(881,113){\usebox{\plotpoint}}
\put(888,113){\usebox{\plotpoint}}
\put(895,113){\usebox{\plotpoint}}
\put(902.0,113.0){\rule[-0.200pt]{0.400pt}{16.140pt}}
\put(909.0,113.0){\rule[-0.200pt]{0.400pt}{16.140pt}}
\put(917.0,113.0){\rule[-0.200pt]{0.400pt}{16.140pt}}
\put(924.0,113.0){\rule[-0.200pt]{0.400pt}{16.140pt}}
\put(931.0,113.0){\rule[-0.200pt]{0.400pt}{16.140pt}}
\put(938.0,113.0){\rule[-0.200pt]{0.400pt}{16.140pt}}
\put(945,113){\usebox{\plotpoint}}
\put(952,113){\usebox{\plotpoint}}
\put(959,113){\usebox{\plotpoint}}
\put(966,113){\usebox{\plotpoint}}
\put(974,113){\usebox{\plotpoint}}
\put(981,113){\usebox{\plotpoint}}
\put(988,113){\usebox{\plotpoint}}
\put(995,113){\usebox{\plotpoint}}
\put(1002,113){\usebox{\plotpoint}}
\put(1009,113){\usebox{\plotpoint}}
\put(1016,113){\usebox{\plotpoint}}
\put(1023,113){\usebox{\plotpoint}}
\put(1031,113){\usebox{\plotpoint}}
\put(1038,113){\usebox{\plotpoint}}
\put(1045,113){\usebox{\plotpoint}}
\put(1052,113){\usebox{\plotpoint}}
\put(1059,113){\usebox{\plotpoint}}
\put(1066,113){\usebox{\plotpoint}}
\put(1073,113){\usebox{\plotpoint}}
\put(1080,113){\usebox{\plotpoint}}
\put(1087.0,113.0){\rule[-0.200pt]{0.400pt}{16.140pt}}
\put(1095.0,113.0){\rule[-0.200pt]{0.400pt}{16.140pt}}
\put(1102.0,113.0){\rule[-0.200pt]{0.400pt}{16.140pt}}
\put(1109.0,113.0){\rule[-0.200pt]{0.400pt}{16.140pt}}
\put(1116.0,113.0){\rule[-0.200pt]{0.400pt}{16.140pt}}
\put(1123.0,113.0){\rule[-0.200pt]{0.400pt}{16.140pt}}
\put(1130,113){\usebox{\plotpoint}}
\put(1137,113){\usebox{\plotpoint}}
\put(1144,113){\usebox{\plotpoint}}
\put(1152,113){\usebox{\plotpoint}}
\put(1159,113){\usebox{\plotpoint}}
\put(1166,113){\usebox{\plotpoint}}
\put(1173,113){\usebox{\plotpoint}}
\put(1180,113){\usebox{\plotpoint}}
\put(1187.0,113.0){\rule[-0.200pt]{0.400pt}{16.140pt}}
\put(1194.0,113.0){\rule[-0.200pt]{0.400pt}{16.140pt}}
\put(1201.0,113.0){\rule[-0.200pt]{0.400pt}{16.140pt}}
\put(1209.0,113.0){\rule[-0.200pt]{0.400pt}{16.140pt}}
\put(1216.0,113.0){\rule[-0.200pt]{0.400pt}{16.140pt}}
\put(1223.0,113.0){\rule[-0.200pt]{0.400pt}{16.140pt}}
\put(1230.0,113.0){\rule[-0.200pt]{0.400pt}{16.140pt}}
\put(1237.0,113.0){\rule[-0.200pt]{0.400pt}{16.140pt}}
\put(1244.0,113.0){\rule[-0.200pt]{0.400pt}{16.140pt}}
\put(1251.0,113.0){\rule[-0.200pt]{0.400pt}{16.140pt}}
\put(1258.0,113.0){\rule[-0.200pt]{0.400pt}{16.140pt}}
\put(1265.0,113.0){\rule[-0.200pt]{0.400pt}{16.140pt}}
\put(1273.0,113.0){\rule[-0.200pt]{0.400pt}{16.140pt}}
\put(1280.0,113.0){\rule[-0.200pt]{0.400pt}{16.140pt}}
\put(1287.0,113.0){\rule[-0.200pt]{0.400pt}{16.140pt}}
\put(1294.0,113.0){\rule[-0.200pt]{0.400pt}{16.140pt}}
\put(1301.0,113.0){\rule[-0.200pt]{0.400pt}{16.140pt}}
\put(1308.0,113.0){\rule[-0.200pt]{0.400pt}{16.140pt}}
\put(1315.0,113.0){\rule[-0.200pt]{0.400pt}{16.140pt}}
\put(1322.0,113.0){\rule[-0.200pt]{0.400pt}{16.140pt}}
\put(1330.0,113.0){\rule[-0.200pt]{0.400pt}{16.140pt}}
\put(1337.0,113.0){\rule[-0.200pt]{0.400pt}{16.140pt}}
\put(1344.0,113.0){\rule[-0.200pt]{0.400pt}{16.140pt}}
\put(1351,113){\usebox{\plotpoint}}
\put(1358,113){\usebox{\plotpoint}}
\put(1365,113){\usebox{\plotpoint}}
\put(1372,113){\usebox{\plotpoint}}
\put(1379,113){\usebox{\plotpoint}}
\put(1387,113){\usebox{\plotpoint}}
\put(1394,113){\usebox{\plotpoint}}
\put(1401,113){\usebox{\plotpoint}}
\put(1408,113){\usebox{\plotpoint}}
\put(1415,113){\usebox{\plotpoint}}
\put(1422,113){\usebox{\plotpoint}}
\put(1429,113){\usebox{\plotpoint}}
\put(1436,113){\usebox{\plotpoint}}
\put(1444,113){\usebox{\plotpoint}}
\put(1451,113){\usebox{\plotpoint}}
\put(1458,113){\usebox{\plotpoint}}
\put(1465,113){\usebox{\plotpoint}}
\put(1472,113){\usebox{\plotpoint}}
\put(1479,113){\usebox{\plotpoint}}
\put(1486,113){\usebox{\plotpoint}}
\put(1493,113){\usebox{\plotpoint}}
\put(1500,113){\usebox{\plotpoint}}
\put(1508.0,113.0){\rule[-0.200pt]{0.400pt}{16.140pt}}
\put(1515.0,113.0){\rule[-0.200pt]{0.400pt}{16.140pt}}
\put(1522.0,113.0){\rule[-0.200pt]{0.400pt}{16.140pt}}
\put(1529.0,113.0){\rule[-0.200pt]{0.400pt}{16.140pt}}
\put(1536.0,113.0){\rule[-0.200pt]{0.400pt}{16.140pt}}
\put(1543.0,113.0){\rule[-0.200pt]{0.400pt}{16.140pt}}
\put(1550.0,113.0){\rule[-0.200pt]{0.400pt}{16.140pt}}
\put(1557.0,113.0){\rule[-0.200pt]{0.400pt}{16.140pt}}
\put(1565.0,113.0){\rule[-0.200pt]{0.400pt}{16.140pt}}
\put(1572.0,113.0){\rule[-0.200pt]{0.400pt}{16.140pt}}
\put(1579.0,113.0){\rule[-0.200pt]{0.400pt}{16.140pt}}
\put(1586.0,113.0){\rule[-0.200pt]{0.400pt}{16.140pt}}
\put(1593.0,113.0){\rule[-0.200pt]{0.400pt}{16.140pt}}
\put(1600.0,113.0){\rule[-0.200pt]{0.400pt}{16.140pt}}
\put(1607.0,113.0){\rule[-0.200pt]{0.400pt}{16.140pt}}
\put(1614.0,113.0){\rule[-0.200pt]{0.400pt}{16.140pt}}
\put(1622.0,113.0){\rule[-0.200pt]{0.400pt}{16.140pt}}
\put(1629.0,113.0){\rule[-0.200pt]{0.400pt}{16.140pt}}
\put(1636.0,113.0){\rule[-0.200pt]{0.400pt}{16.140pt}}
\put(1643.0,113.0){\rule[-0.200pt]{0.400pt}{16.140pt}}
\put(1650.0,113.0){\rule[-0.200pt]{0.400pt}{16.140pt}}
\put(1657,113){\usebox{\plotpoint}}
\put(1664,113){\usebox{\plotpoint}}
\put(1671,113){\usebox{\plotpoint}}
\put(1678,113){\usebox{\plotpoint}}
\put(1686,113){\usebox{\plotpoint}}
\put(1693,113){\usebox{\plotpoint}}
\put(1700,113){\usebox{\plotpoint}}
\put(1707,113){\usebox{\plotpoint}}
\put(1714,113){\usebox{\plotpoint}}
\put(1721,113){\usebox{\plotpoint}}
\put(1728,113){\usebox{\plotpoint}}
\put(1735,113){\usebox{\plotpoint}}
\put(1743,113){\usebox{\plotpoint}}
\put(1750,113){\usebox{\plotpoint}}
\put(1757,113){\usebox{\plotpoint}}
\put(1764,113){\usebox{\plotpoint}}
\put(1771,113){\usebox{\plotpoint}}
\put(1778,113){\usebox{\plotpoint}}
\put(1785,113){\usebox{\plotpoint}}
\put(1792,113){\usebox{\plotpoint}}
\put(1800,113){\usebox{\plotpoint}}
\put(1807,113){\usebox{\plotpoint}}
\put(1814,113){\usebox{\plotpoint}}
\put(1821,113){\usebox{\plotpoint}}
\put(1828,113){\usebox{\plotpoint}}
\put(1835,113){\usebox{\plotpoint}}
\put(1842,113){\usebox{\plotpoint}}
\put(1849,113){\usebox{\plotpoint}}
\put(1857,113){\usebox{\plotpoint}}
\put(1864,113){\usebox{\plotpoint}}
\put(1871,113){\usebox{\plotpoint}}
\put(1878,113){\usebox{\plotpoint}}
\put(1885,113){\usebox{\plotpoint}}
\end{picture}

\vspace*{0.5cm}
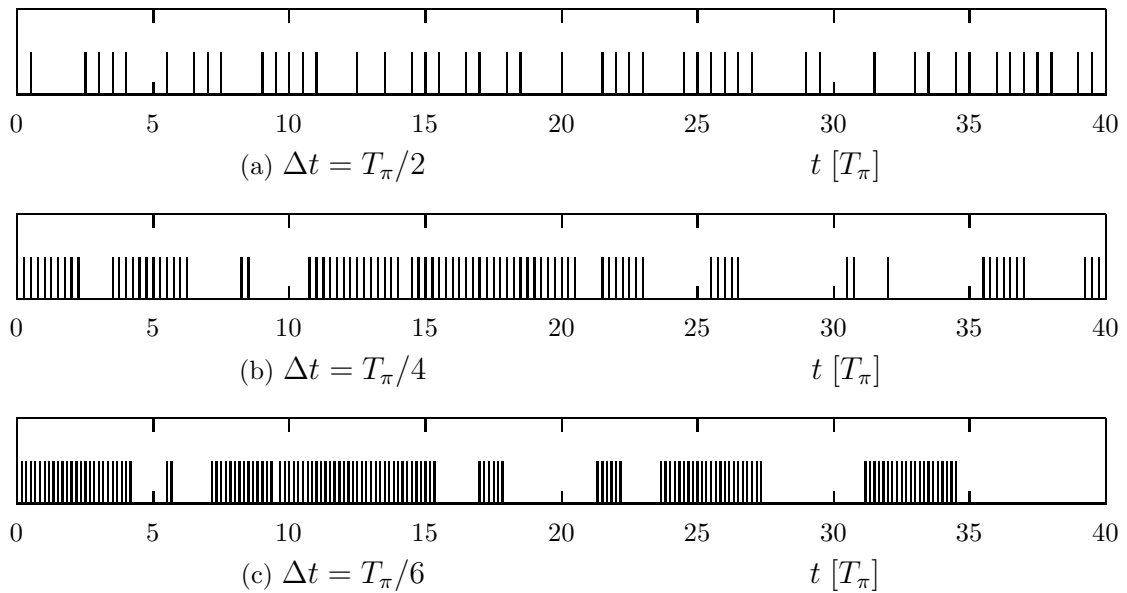
Fig. 2. Stochastic alternating light and dark periods. The lines
mark times when the atom is found in state $|1\rangle$ and emits a
burst of light. $T_\pi=\pi/\Omega_2$ is the length of a $\pi$ pulse.

\end{document}